\documentclass[12pt]{iopart}
\usepackage{iopams}
\usepackage{color}
\usepackage{graphicx}
\usepackage{cite}
\begin{document}

\title{Transport and interaction blockade of cold bosonic atoms in a triple-well potential}

\author{P. Schlagheck$^{1,2}$, F. Malet$^1$, J. C. Cremon$^1$, and S. M. Reimann$^1$}

\address{$^1$ Mathematical Physics, LTH, Lund University, Box 118, 22100 Lund, Sweden}
\address{$^2$ D\'epartement de Physique, Universit\'e de Li\`ege, 4000 Li\`ege, Belgium}

\begin{abstract}

We theoretically investigate the transport properties of cold bosonic atoms
in a quasi one-dimensional 
triple-well potential that consists of two large outer wells, which act
as microscopic source and drain reservoirs, and a small inner well, which
represents a quantum-dot-like scattering region. Bias and gate ``voltages''
introduce a time-dependent tilt of
the triple-well configuration, and are used 
to shift the energetic level of the inner
well with respect to the outer ones.
By means of exact diagonalization considering a total number of six atoms in
the triple-well potential, we find diamond-like structures for the
occurrence of single-atom transport in the parameter space spanned by the
bias and gate voltages.
We discuss the analogy with Coulomb blockade in electronic quantum dots,
and point out how one can infer the interaction energy in the central well
from the distance between the diamonds.

\end{abstract}


\maketitle

\section{Introduction}

The development of quasi one-dimensional waveguides for cold atoms based on optical
lattices \cite{MorObe06RMP} and atom chips \cite{FolO00PRL,ForZim07RMP} has lead to
a number of theoretical investigations on the guided quasi-stationary and dynamical
transport properties of ultracold atomic gases
\cite{LebPav01PRA,LebPavSin03PRA,CarLar00PRL,Car01PRA,PauRicSch05PRL,PauO07PRA}.
Particular attention was devoted to the transport of coherent bosonic matter waves through
quantum-dot potentials realized, e.g., by two (magnetic or optical) barriers along
the waveguide, which display nonlinear transmission features that are reminiscent from
nonlinear optics \cite{CarLar00PRL,Car01PRA,PauRicSch05PRL,PauO07PRA,RapWitKor06PRA}.
An important long-term aim in this context is to establish a close analogy with
\emph{electronic} conduction through microfabricated structures and nanostructures
in solid-state systems.
This latter field of research still exhibits a number of open questions especially
related to the quantitative role of interaction and correlation between the electrons
in such mesoscopic transport processes.

In spite of remarkable advancements in the creation of one-dimensional matter-wave
beams through guided atom lasers \cite{GueO06PRL,CouO08EPL}, the experimental
realization and evaluation of such one-dimensional atomic scattering processes still
represents a formidable task.
It makes therefore sense to consider alternative approaches based on \emph{closed}
systems, where two reservoirs of ultracold atoms are connected to each other and
biased such that a flow of atoms can take place from one reservoir to the other.
Atomic quantum dots across which the atoms have to tunnel could be implemented
by inducing small potential wells in between the reservoirs.
Such geometries were studied in recent investigations on transistor-like operations
with Bose-Einstein condensates \cite{StiAndZoz07PRA}.
The quantification of ``conduction'' across such a quantum dot under a small bias,
however, requires a precise measurement of atomic populations ideally on a
single-atom level, which is not impossible \cite{CamO06PRA} but seems rather
hard to be realized in this specific context.

A solution to the problem of counting individual atoms was recently provided by
experiments with optical double-well lattices, focusing on correlated two-particle
tunneling \cite{FoeO07N} as well as on the 
interaction blockade~\cite{CapO07PRL} with three particles
\cite{CheO08PRL}.
In these experiments, each double-well site was identically prepared with a well-defined
number of atoms and finally ``read out'' by absorption imaging after time-of-flight
expansion, where the populations in the left and right wells were, before switching
off the lattices, transferred to different Brillouin zones.
This parallel processing of identical few-particle experiments especially allowed for
the detection of integer atomic populations in each of the two wells per site
\cite{CheO08PRL}, as a consequence of the strong repulsive interaction between
the atoms. More recently, much theoretical work has focused on the tunneling
of bosons in such double well potentials, see for 
example \cite{dounas2007,zollner2007,zollner2008a,zollner2008b,averin2008}.

Inspired by the basic idea sketched above, 
we now propose here to investigate source-drain transport
processes with ultracold atoms on the basis of optical \emph{triple-well} lattices,
which could possibly be realized by adding another standing-wave beam with the appropriate
wavelength to the experimental setup in Refs.~\cite{FoeO07N,CheO08PRL}.
If all three lattice potentials that together form the triple-well lattice are imposed
with about the same amplitude and with a suitable phase shift relative to each other,
a triple-well potential can be created on each site where the two outer wells are
considerably larger than the inner one.
After loading the lattice with a well-defined number of atoms per site, these outer wells
can be regarded as microscopic ``source'' and ``drain'' reservoirs,
while the central well acts as an atomic quantum dot.

We now use this general configuration in this paper in order to study the atomic analog
of \emph{Coulomb blockade} in electron transport 
(see for example \cite{LamJak69PRL,averin1986,averin1991,grabert1991,meir1991} 
and the review \cite{ReiMan02RMP}) in the conduction of strongly
interacting bosonic atoms from the source to the drain across the dot region.
We consider, to this end, a time-dependent ramping process of a bias between the outer wells,
which can be implemented by varying the amplitude and phase of the main (longest-wavelength)
component of the triple-well lattice, and address the question to which extent this ramping
process leads to the transfer of one or several atoms from the source reservoir to the dot
or from there to the drain reservoir.
While this perturbation represents the analog of a ``bias voltage'' in electronic quantum
dots, a ``gate voltage'', which lowers the on-site eigenenergies of the central well
relative to the outer ones, can be induced in a very similar manner, again by a suitable
manipulation of the main laser beam of the lattice.
Our aim is to map out the lines of finite ``conductance'', i.e.\ of a finite transfer of
atoms between the wells, in the parameter space spanned by the above-mentioned gate and bias voltages.
We shall show that this gives rise to diamond-like structures that are closely analogous
to Coulomb diamonds in electronic quantum dots.

We start in Section \ref{sec:system} with a detailed description of the triple-well system
under consideration, which is defined in accordance with the experimental setups used in
Refs.~\cite{FoeO07N,CheO08PRL}.
Section \ref{sec:static} is devoted to the discussion of the ``static'' interaction blockade
in the triple-well system in perfect analogy with the double-well interaction blockade
experiment in Ref.~\cite{CheO08PRL}.
In Section \ref{sec:transport}, we discuss the outcome of the time-dependent ramping process,
which is numerically computed both through diagonalization of the many-body Hamiltonian as
well as through the propagation of the many-body wavefunction under the variation of the bias.
The resulting diamond structures are explained within a simple Bose-Hubbard model and compared
with electronic Coulomb diamonds.
We finally discuss the relevant energy scales of this transport process and point out possible
implications for future research in the Conclusion.

\section{The setup}

\label{sec:system}

\begin{figure}[t]
\centerline{\includegraphics[width=\linewidth]{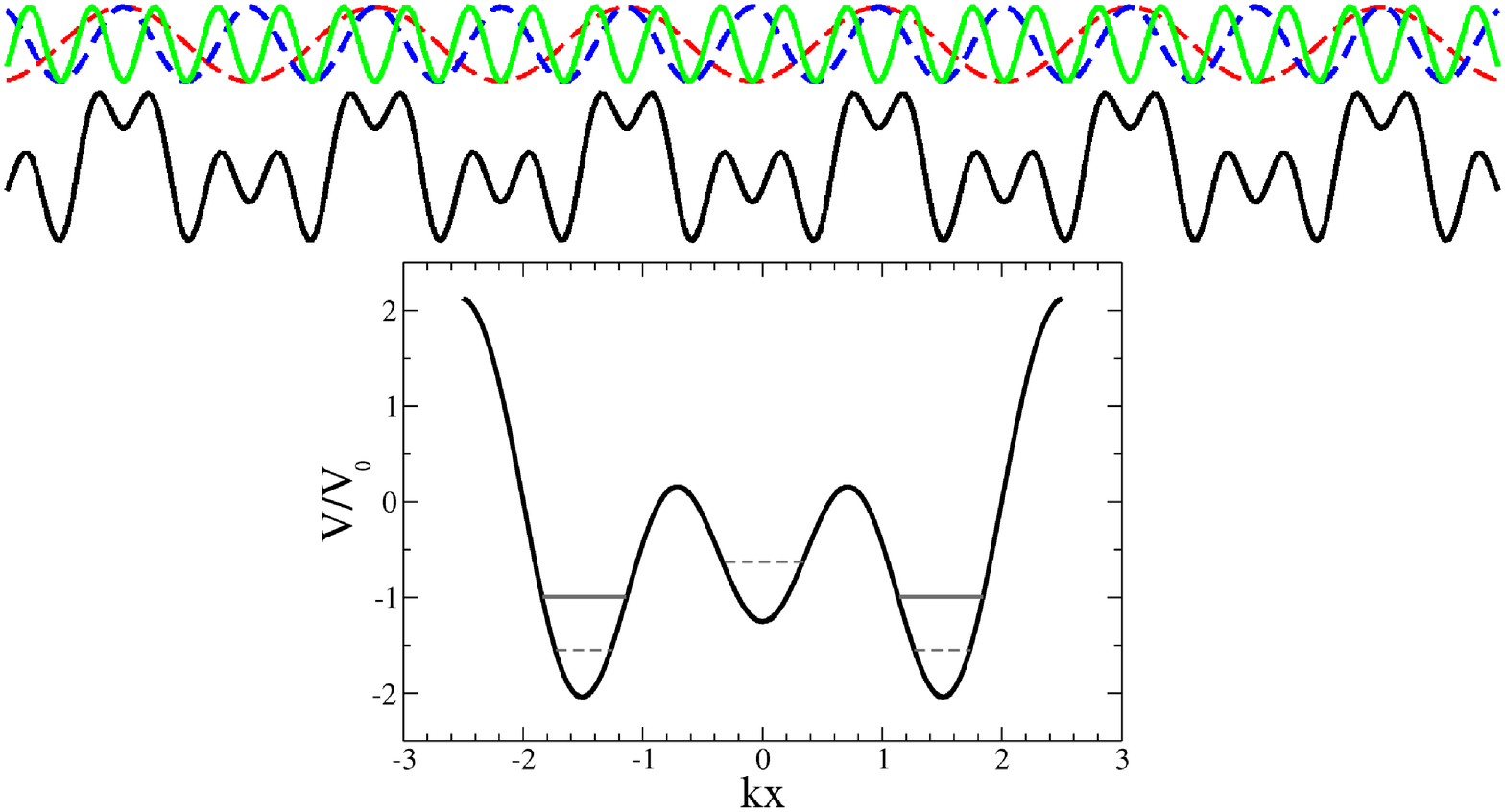}}
\caption{(color online)
Three optical lattices with the commensurate periods $2 \pi / k$,  $4 \pi / k$,
and $8 \pi / k$ (upper panel) are superimposed according to Eq.~(\ref{eq:potential})
to form a superlattice (middle panel).
The lower panel displays the resulting triple-well potential on an individual site of
the superlattice (for $V_g = V_b = 0$).
The dashed lines mark the unperturbed single-particle energies in the wells, and the
solid lines indicate the chemical potential, or, more precisely, the particle-removal
energy of an individual atom for the six-particle ground state where three atoms are
localized in the left as well as in the right well.
\label{fig:potential}}
\end{figure}

We consider a gas of ultracold $^{87}$Rb atoms in a quasi one-dimensional confinement
that is exposed to a periodic potential of the form
\begin{equation} \label{eq:potential}
  V(x) = V_0(-\cos kx + \cos 2kx - \cos 4kx) - V_g \cos kx - V_b \sin kx \;.
\end{equation}
This potential can be generated by a superposition of three counter-propagating
laser beams with the wavelengths $\lambda_1 = \pi / k$, $\lambda_2 = 2 \pi / k$,
and $\lambda_3 = 4 \pi / k$.
We specifically consider the wavelengths $\lambda_1 = 1530$ nm and $\lambda_2 = 765$ nm,
which were also used in the experiments of Refs.~\cite{FoeO07N,CheO08PRL}, and assume for the
third laser beam the wavelength $\lambda_3 = 382.5$ nm.
The latter could possibly be realized by a frequency-doubling of the laser beam with
the wavelength $\lambda_2$ or, alternatively, by retro-reflecting a part of this laser beam
(split with an acousto-optical deflector) under a finite angle of $60^\circ$ as done in
Ref.~\cite{KieO08PRL}.
Using suitable phase shifts between these three counterpropagating laser beams (and taking into
account the fact that the wavelength $\lambda_1$ is red-detuned while the wavelengths
$\lambda_2$, $\lambda_3$ are blue-detuned with respect to the intra-atomic transition
in $^{87}$Rb), we thereby obtain the effective potential
$V_{\rm eff}(x) = - 2 V_1 \cos^2(\frac{1}{2} k x - \phi / 2) + 2 V_2 \cos^2( k x )
  + 2 V_3 \cos^2( 2 k x - \pi / 2 )$
with positive prefactors $V_1$, $V_2$, and $V_3$, which apart from a constant offset is
exactly equivalent to Eq.~(\ref{eq:potential}) provided we choose $V_1 \cos \phi = V_0 + V_g$,
$V_1 \sin \phi = V_b$, and $V_2 = V_3 = V_0$.
In this way, we obtain, as shown in Fig.~\ref{fig:potential}, a periodic lattice of sites with
triple-well potentials, where on each site the two outer wells are deeper than the central well.

The confinement to one spatial direction is assumed to be ensured by the presence of a strong
two-dimensional (red-detuned) optical lattice in the transverse spatial directions, described
by the effective potential $V_{\rm tr}(y,z) = - V_\perp ( \cos k_\perp y + \cos k_\perp z )^2$.
This gives rise to a lattice of harmonic waveguides parallel to the $x$-axis with the
confinement frequency $\omega_\perp = 2 k_\perp \sqrt{V_\perp / m}$ where $m$ denotes the mass
of a $^{87}$Rb atom.
As was shown by Olshanii \cite{Ols98PRL}, the effective one-dimensional interaction strength
along these waveguides is given by
\begin{equation} \label{eq:g1D}
  g = \frac{2 \hbar \omega_\perp a_s}{1 - C a_s / a_\perp}
\end{equation}
with $C \simeq 1.4603$, where $a_s \simeq 5.8$ nm is the $s$-wave scattering length for
$^{87}$Rb atoms and $a_\perp \equiv \sqrt{\hbar/(m \omega_\perp)}$ denotes the transverse
oscillator length.
Obviously, a rather large value for $V_\perp$ is generally required in order to induce
strong interactions along the one-dimensional confinement.
Setting, as in the experiments of Refs.~\cite{FoeO07N,CheO08PRL}, the wavelength of the
transverse laser beams to $\lambda_\perp \equiv 2 \pi / k_\perp = 843$ nm, we obtain the
one-dimensional interaction strength $g = 4 \hbar^2 k / m$ if we choose
$V_\perp = 690 E_r^\perp$ with $E_r^\perp \equiv \hbar^2 k_\perp^2 / (2 m)$ being the
transverse recoil energy.

We shall in the following consider also a large amplitude of the longitudinal triple-well
lattice and set, for the sake of definiteness, the prefactor $V_0$ in Eq.~(\ref{eq:potential})
to $V_0 = 20 \hbar^2 k^2 / m = 40 E_r$ where $E_r \equiv \hbar^2 k^2 / (2 m)$ (corresponding
to $\simeq 7.8$ kHz) is the recoil energy of the laser with the wavelength $\lambda_2$.
For this particular choice of the lattice strength, and for $V_b = V_g = 0$, the energetically
lowest single-particle eigenstates within each triple-well site are strongly localized in the
wells.
A rather small splitting $\delta E_{LR} \simeq 0.02 \hbar^2 k^2 / m$ of the ground-state doublet,
consisting of the positive and negative linear combination of the ``Wannier states''
(or quasi-modes) in which the atom is localized around the center of the left and right well,
respectively, is thereby obtained, corresponding to a tunneling time scale of the order of
$\tau \sim \hbar / \delta E_{LR} \simeq 1$ms.
This splitting is much smaller than the level spacing between the ground-state doublet and
the doublet containing an excited state in one of the wells, which is roughly characterized
by the local harmonic-oscillator energy
$\hbar \omega_{||} \simeq 4 \hbar k \sqrt{V_0 / m} \simeq 18 \hbar^2 k^2 / m$
within the well.
Tunneling between \emph{adjacent} triple-well sites is neglected in the following, as the
associated inter-site tunneling time exceeds all other relevant time scales of this system.
We therefore consider the dynamics within the individual triple-well sites to be completely
independent of each other.

In addition to the ``main'' triple-well lattice characterized by the amplitude $V_0$,
we introduce in Eq.~(\ref{eq:potential}) two independent perturbations with the
amplitudes $V_g$ and $V_b$ which can be controlled by a suitable manipulation of the
laser beam with the largest wavelength $\lambda_1$.
In close analogy with Coulomb blockade experiments in electronic quantum dots, we name
$V_g$ the ``gate voltage'' and $V_b$ the ``bias voltage''.
Indeed, as for electronic quantum dots, the effect of increasing $V_g$ is to lower the
energetic offset of the central well with respect to the outer ones, which allows one to
enhance the ground-state population of this well in the many-body system.
A positive bias voltage $V_b$, on the other hand, gives rise to an overall tilt of the
triple-well configuration, which opens the possibility for a transfer of individual atoms
from the left to the central or from the central to the right well.

In lowest order, the impact of the gate and bias voltages $V_g$ and $V_b$ can approximately
be quantified by the shifts $E_L \to E_L + V_b$, $E_C \to E_C - V_g$, and $E_R \to E_R - V_b$,
of the single-particle energies $E_L$, $E_C$, and $E_R$ in the left, central, and right well,
respectively.
This result is obtained by a simple perturbative consideration where the matrix elements of
the gate and bias perturbations in Eq.~(\ref{eq:potential}) are evaluated within the local
ground states in the wells, which are approximated by Gaussian wavefunctions centred
around $kx = 0$ for the central and around $kx = \pm \pi / 2$ for the right and left well.
Using $\omega_{||} \simeq 4 k \sqrt{V_0 / m}$ as local oscillator frequency and
$V_0 = 20 \hbar^2 k^2 / m$, we obtain $E_{L/R} \to E_{L/R} \pm 0.986 V_b$ 
and $E_C \to E_C - 0.986 V_g$.
This implies that the gate and bias ``voltages'' directly correspond here to the associated
energy shifts in the individual wells.

\section{Interaction blockade in the ground-state populations}

\label{sec:static}

As in Ref.~\cite{CheO08PRL}, we assume that the superlattice is initially loaded with an
ultracold gas of $^{87}$Rb atoms, such that each of the triple-well sites is populated
with a given well-defined number of atoms.
In the experiment of Ref.~\cite{CheO08PRL} which used a double-well lattice, interaction
blockade was measured by detecting the ground-state populations in the left and right wells
as a function of a finite bias between the wells.
For this purpose, the population in the left well was transferred into a highly excited
state within the double-well potential by means of a rather fast increase of the bias;
this allows to separately detect it through time-of-flight expansion and subsequent absorption
imaging after switching off the optical lattices (see also Ref.~\cite{FoeO07N}).
A step-like structure, with plateaus at integer values, was found for the population in the
left and right wells as a function of the bias \cite{CheO08PRL}, which is a clear
consequence of the strong repulsive interaction between the atoms.

\begin{figure}[t]
\centerline{\includegraphics[width=\linewidth]{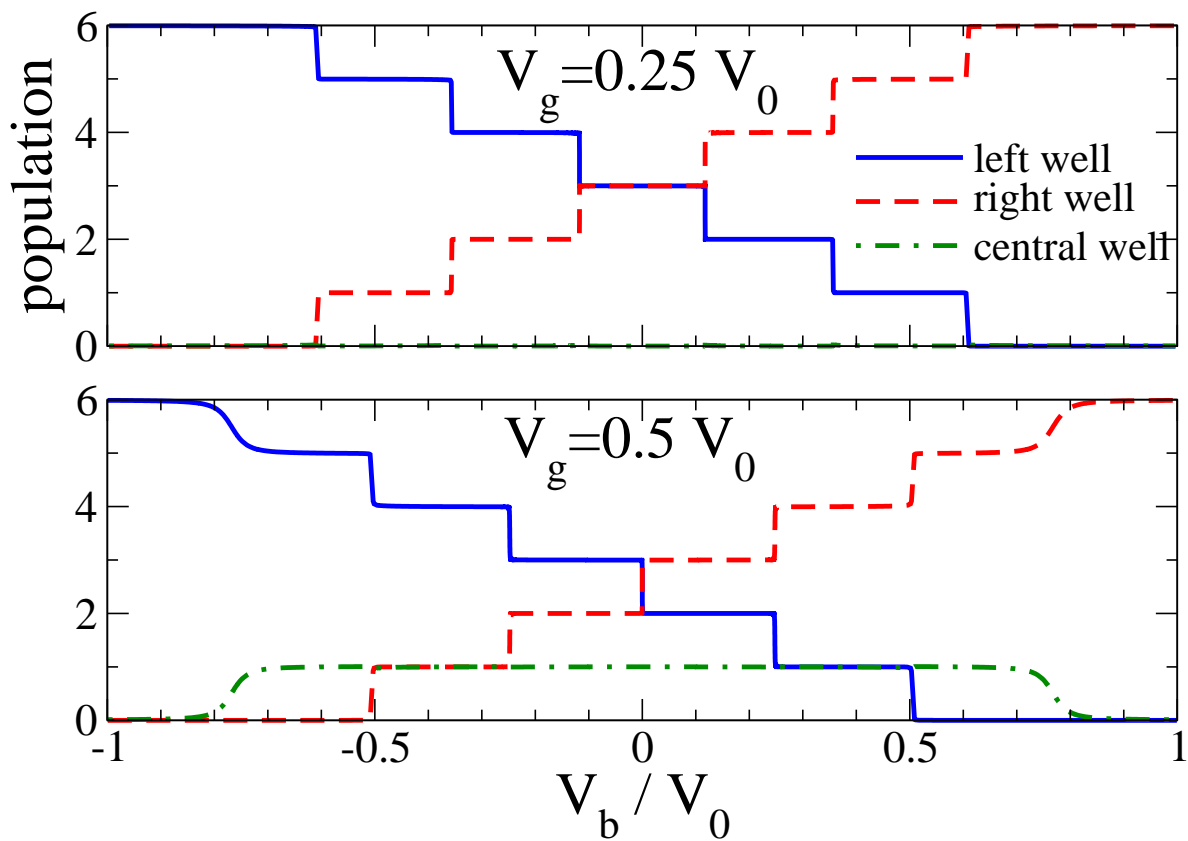}}
\caption{(color online)
Populations in the left (solid blue line), right (dashed red line), and central well
(dot-dashed green line) for the exact ground state of six particles in the triple-well
potential as a function of the bias voltage $V_b$.
The calculation was done for the gate voltages $V_g = 0.25 V_0$ (upper panel), for which
the central well is practically empty, and for $V_g = 0.5 V_0$ (lower panel), for which the
central well contains one atom at not too strong bias.
Due to the strong repulsive interaction between the atoms, the populations in the wells
undergo step-like transitions between plateaus at integer values under the variation
of the potential, in close analogy with interaction blockade in double-well lattices
\cite{CheO08PRL}.
\label{fig:groundstates}}
\end{figure}

This programme can be carried out as well for our triple-well configuration.
Figure \ref{fig:groundstates} shows, as a function of the bias voltage $V_b$,
the populations in the left, central, and right well that is contained in the many-body
ground state of six particles per triple-well site.
As in Ref.~\cite{CheO08PRL}, the populations undergo rather sharp, step-like transitions
between plateaus at integer values, which clearly underlines the relevance of the strong
repulsive interaction between the atoms.
While the central well remains practically empty at the gate voltage $V_g = 5$, it is
populated with one atom at $V_g = 10$ provided the bias is not too strong.
This scenario can straightforwardly be generalized to stronger gate voltages that would
allow for two or more atoms in the central well.
Decreasing the interaction strength will lead to a decrease of the bias voltage interval
in which the sequence of transitions takes place, until in the limit of vanishing
interaction this sequence shrinks down to a collective transition from a fully left-biased
state, with all atoms in the left well, at negative to a fully right-biased state at positive
bias voltages.
An infinitely large interaction strength $g$ will, on the other hand, not give rise to an
infinite interaction energy within the wells and to an infinite extent of the plateaus,
but rather approach the case of noninteracting spinless \emph{fermionic} atoms
\cite{Ols98PRL} where the Pauli exclusion principle is responsible for the appearance of
plateaus and step-like transitions in the populations.

The many-body ground states were calculated by means of an exact
diagonalization approach based on the Lanczos algorithm \cite{Lan50JRNBS}, which takes
into account all Fock states with a given total number of particles that are defined
upon a suitably truncated single-particle basis.
We chose as basis functions the plane waves $\phi_n(x) = \sqrt{k / (2 \pi)} e^{i n k x}$
that satisfy periodic boundary conditions within the triple-well lattice, and took into
account for the many-body calculation all $\phi_n$ with $-10 \leq n \leq 10$.
We verified that for six particles this truncation is sufficient to reproduce all relevant
features of the many-body states under consideration.
After calculating the ground state and its many-body eigenfunction, the populations in the
individual wells are computed by integrating the spatial atomic density in between
(artificial) separation points at the local maxima of the triple-well potential.
The population in the left well is thus obtained as the integral over the density from
$x = - \pi / k$ to the position of the left local maximum (at $x \simeq - 0.7 / k$ for
$V_g = V_b = 0$), while the population in the central well is given by the integral
over the density in between the left and right maxima.

With few exceptions (occuring notably at near-degeneracies between the energies of
different many-body states), we almost always obtain nearly integer well populations
for the eigenstates, due to the strong repulsive interaction between the atoms.
We shall therefore denote these eigenstates by $N_L$:$N_C$:$N_R$ in the following,
where the integers $N_L$, $N_C$, and $N_R$ represent the populations in the left,
central, and right well, respectively.

\section{Interaction blockade in the transport}

\label{sec:transport}

\subsection{Time-dependent ramping process}

In contrast to the double-well potential, the confinement (\ref{eq:potential}) permits
to probe interaction blockade not only in the ``static'' properties of the many-body
ground state, but also in the dynamical \emph{transport} behaviour of the system
under the variation of the bias.
We assume for this purpose that the triple-well lattice is initially loaded with a
given number of atoms at a given gate voltage $V_g$ and at vanishing bias $V_b = 0$.
The bias voltage is then \emph{dynamically ramped} from zero to a given maximal value
on a suitable time scale.
After this ramping process, the population in the individual wells is measured by a
suitable transfer of the atoms to excited modes, time-of-flight expansion, and absorption
imaging \cite{FoeO07N}.

The speed of this ramping has to be chosen such that is is slow with respect to the
tunneling time scale between adjacent wells (i.e., between the left and central well,
or between the central and right well), but fast with respect to ``direct'' tunneling
between the left and right well.
More quantitatively, considering a linear ramping process of the form
$V_b(t) = s t$ with constant speed $s$, and taking into account that the shift of the
single-particle energy in the left or right well with respect to the central well is
approximately given by $V_b$ (as pointed out in Section \ref{sec:system}), the
Landau-Zener theory for nonadiabatic transitions \cite{Zen32PRSL} predicts the
probability
\begin{equation}
  P = 1 - \exp[ - 2 \pi \Delta E^2 / (\hbar s) ] \label{eq:LZ}
\end{equation}
for the transfer of an atom at an avoided crossing between the state $N_L$:$N_C$:$N_R$
and the state $N_L\pm 1$:$N_C \mp 1$:$N_R$ (or $N_L$:$N_C \pm 1$:$N_R \mp 1$),
where $\Delta E$ denotes the level splitting between the hybridized states right
at the anticrossing.
If we denote by $\delta E$ the analogous level splitting between $N_L$:$N_C$:$N_R$
and $N_L\pm 1$:$N_C$:$N_R \mp 1$, corresponding to the transfer of an atom across
both barriers, we obtain the requirement that the speed $s$ of the ramping process
would have to satisfy $\delta E^2 \ll \hbar s < \Delta E^2$.

\begin{figure}[t]
\centerline{\includegraphics[width=\linewidth]{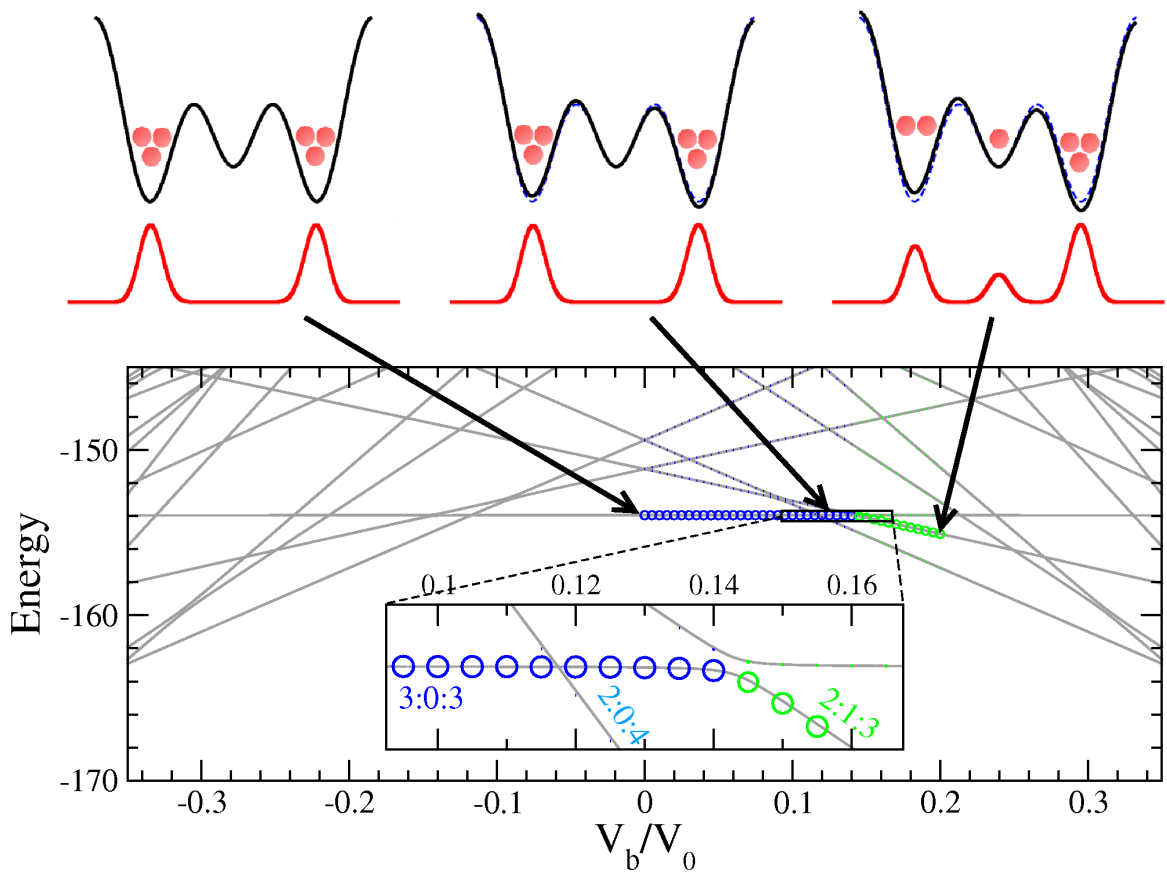}}
\caption{(color online)
Time-dependent transport process for six particles in the triple-well potential.
We start here from an unbiased configuration at gate voltage $V_g = 0.25 V_0$
(upper left panel), which is then exposed to a time-dependent sweep of the bias voltage
from $V_b = 0$ to $V_b = 0.2 V_0$ with the constant speed 
$d V_b / d t = 0.002 \hbar^3 k^4 / m^2$.
The upper panels show the potential (above, in black) and the spatial atom densities
(below, in red) for the bias voltages $V_b = 0$ (left panel), $V_b = 0.125 V_0$
(middle panel), and $V_b = 0.2 V_0$ (right panel).
The lower panel illustrates the evolution of this time-dependent ramping process
in the many-body spectrum (energies are in units of $\hbar^2 k^2 / m$).
We compute for this purpose the decomposition of the time-dependent many-body
wavefunction into the instantaneous eigenstates of the Hamiltonian, and represent
the relative weight of this decomposition by the size of the circles.
Clearly, the system undergoes a nearly perfect transition from the 3:0:3 to the 2:1:3
state at the avoided crossing at $V_b \simeq 0.143 V_0$, and is nearly not affected
by the previous anticrossing between the 3:0:3 and 2:0:4 states at $V_b \simeq 0.117 V_0$.
\label{fig:tilt}}
\end{figure}

In the spirit of this consideration, the outcome of this time-dependent ramping
process can be predicted by computing not only the ground state, but also lowly
excited states of the many-body system as a function of the bias voltage.
The lower panel in Fig.~\ref{fig:tilt} shows the result of such a numerical
calculation, which was done for six particles at the gate voltage $V_g = 0.25 V_0$.
In accordance with Fig.~\ref{fig:groundstates}, we recognize that the state with
the populations 3:0:3, which is the energetically lowest state at zero bias,
undergoes a small anticrossing with the 2:0:4 state at $V_b \simeq 0.117 V_0$,
which then represents the new ground state beyond that value of the bias voltage.
The level splitting at this anticrossing is found to be
$\delta E \simeq 0.001 \hbar^2 k^2 / m$,
which is much smaller than the subsequent splitting
$\Delta E \simeq 0.069 \hbar^2 k^2 / m$
between the levels of the 3:0:3 and the 2:1:3 states.
Consequently, a ramping process $V(t) = s t$ with, e.g., the constant speed
$s = 0.002 \hbar^3 k^4 / m^2$, altogether taking place within a time interval
of $\tau \simeq 0.2 V_0 / s \simeq 40$ ms,
will provide the desired atom transfer from the left to the central well.

This is indeed confirmed by a truly time-dependent simulation of the ramping process.
We use for this purpose a variant of the Lanczos algorithm \cite{ParLig86JCP},
which involves the creation of a finite Krylov subspace that contains the most
relevant components describing the time derivative of the wavefunction to be
propagated.
Within short time intervals $\delta t$, a numerically precise propagation is
carried out in this Krylov subspace, utilizing the representation of both the
unperturbed part of the Hamiltonian and the time-dependent perturbation within
this subspace \cite{note}.
In between those time intervals, the Krylov subspace is recreated using the ``new''
wavefunction that results from this propagation step.
In practice, we find good convergence using 100 Krylov vectors and the time step
$\delta t = 1.0 m / (\hbar k^2)$.

The lower panel of Fig.~\ref{fig:tilt} shows the result of such a time-dependent
calculation for a ramping process of the bias voltage with the speed
$s = 0.002 \hbar^3 k^4 / m^2$ from $V_b = 0$ to $V_b = 0.2 V_0$.
To illustrate the evolution in the many-body spectrum, we compute here
at regular intervals in time the overlap matrix elements of the wavefunction
with the instantaneous eigenstates of the Hamiltonian.
This decomposition is then represented in Fig.~\ref{fig:tilt} by circles whose
sizes are directly proportional to the square moduli of those matrix elements.
We basically see that the time-dependent wavefunction is, at almost all
considered bias voltages, essentially described by one single eigenstate
of the Hamiltonian, namely the one that exhibits the populations 3:0:3 before
and 2:1:3 after the avoided crossing at $V_b \simeq 0.143 V_0$.
The small anticrossing between the 3:0:3 and the 2:0:4 states at
$V_b \simeq 0.117 V_0$, on the other hand, is almost completely ``ignored''
by the time-dependent wavefunction, which undergoes a \emph{diabatic} transition
across this anticrossing (in contrast to the adiabatic transition at
$V_b \simeq 0.143 V_0$).

\begin{figure}[t]
\centerline{\includegraphics[width=\linewidth]{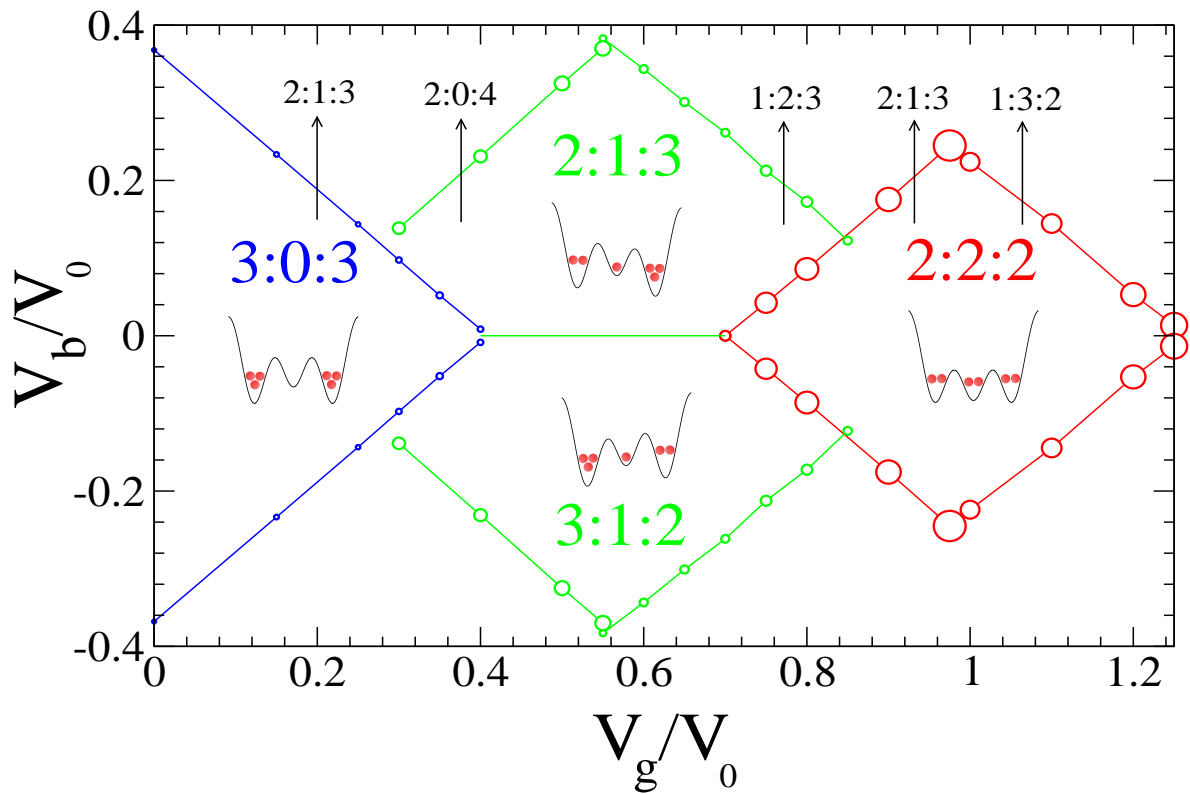}}
\caption{(color online)
  Interaction-blockade diamonds for six atoms in the triple-well potential.
  Plotted are the minimal (positive and negative) bias voltages $V_b$ until
  which the time-dependent ramping process (symbolized by the vertical arrrows)
  has to be performed in order to achieve single-atom transfer between adjacent wells.
  The circles mark the gate voltages for which numerical calculations, based
  on the computation of the many-body spectrum, were explicitly carried out.
  The size of the circles indicate the size $\Delta V_b$ of the avoided crossings
  according to Eq.~(\ref{eq:crossing}), which is proportional to their level splitting.
  Note that the avoided crossings become larger with larger $V_g$, which corresponds
  to the fact that the barrier height between the outer wells and the central well
  is lowered by increasing the gate voltage.
\label{fig:diamonds}}
\end{figure}

This calculation can be carried out for other gate voltages as well.
Figure \ref{fig:diamonds} shows, as a function of $V_g$, the minimal
(positive and negative) bias voltages $V_b$ until which the time-dependent ramping
process has to be performed in order to obtain single-atom transfer between adjacent wells.
The starting configuration depends on the gate voltage under consideration.
For $V_g < 0.4 V_0$, we start with a 3:0:3 population at zero bias, which is then
transformed into a 2:1:3 state under the increase of $V_b$, and into a 3:1:2 state
if instead the bias is ramped into the opposite direction leading to negative $V_b$.
Within $0.4 V_0 < V_g < 0.72 V_0$, the initial state either corresponds to a 2:1:3 or
to a 3:1:2 population, depending on the presence of a tiny positive or negative bias.
The final state of the ramping process depends now on the gate voltage.
For $V_g < 0.56 V_0$, we obtain the 2:0:4 state (or the 4:0:2 state for negative
bias), i.e.\ the atom in the central well is released to the right (left) well, while
for $V_g > 0.56 V_0$ another atom from the left (right) well is pulled into the central
well leading to the 1:2:3 (3:2:1) state.
A completely analogous situation is realized for gate voltages within
$0.72 V_0 < V_g < 1.28 V_0$ where we initially encounter a 2:2:2 population ---
with the only exception that this population does obviously not depend on the presence
of a small initial bias.

\subsection{Bose-Hubbard theory of the diamonds}

The structure and shape of the transitions lines in the $V_b$--$V_g$ parameter space
are strongly reminiscent of Coulomb diamonds in electronic quantum dots (see
for example, the very recent work of~\cite{fuhrer2007} 
or the review \cite{ReiMan02RMP}).
This analogy can indeed be straightforwardly worked out in terms of a simple
Bose-Hubbard-type model.
We take into account for this purpose the three energetically lowest single-particle
eigenfunctions (or rather Wannier functions in the case of degeneracies)
$\phi_L(x)$, $\phi_C(x)$, and $\phi_R(x)$,
corresponding to the particle being localized in the left, central, and right well.
Approximating these single-particle eigenfunctions by normalized Gaussians centred
around the minima of the $(- \cos 4kx)$-component of the potential, with a width that
corresponds to the local oscillator length $a_{||} = \sqrt{\hbar/(m \omega_{||})}$ with
$\omega_{||} \simeq 4 k \sqrt{V_0 / m}$, the shift of the corresponding single-particle
energies due to the presence of finite gate and bias voltages is directly given by
$V_g$ and $V_b$, respectively.
More precisely, these energies are approximately given by $E_L = E_0 + V_b$,
$E_C = E_1 - V_g$, and $E_R = E_0 - V_b$, with
$E_0 \equiv \langle \phi_{L/R}| H_0 | \phi_{L/R} \rangle$ and
$E_1 \equiv \langle \phi_C| H_0 | \phi_C \rangle$,
where we define
$H_0  \equiv  - \frac{\hbar^2}{2 m} \frac{\partial^2}{\partial x^2}
  + V_0(-\cos kx + \cos 2kx - \cos 4kx)$.

The contact interaction $U(x_1 - x_2) = g \delta( x_1 - x_2 )$ gives rise
to the local interaction energies
$E_U^{(L/C/R)} = g \int |\phi_{L/C/R}(x)|^4 dx$ within each well.
Using again the above Gaussian ansatz for the single-particle eigenfunctions,
these interaction energies are approximately equal to each other and, for
$V_0 = 20 \hbar^2 k^2/m$, given by
\begin{equation}
  E_U^{(L/C/R)} \simeq g \sqrt{\frac{m \omega_{||}}{2 \pi \hbar}} \simeq 5.37
  \frac{\hbar^2 k^2}{m} \simeq 0.268 V_0 \equiv E_U . \label{eq:EU}
\end{equation}
Neglecting tunnel couplings between the wells, we obtain the Fock states
$|N_L,N_C,N_R\rangle$ built upon the single-particle basis
$(\phi_L,\phi_C,\phi_R)$ as eigenstates of the many-body system, where
$N_L,N_C,N_R$ represent the populations in the left, central, and right well,
respectively.
The corresponding eigenenergies read
\begin{equation}
  E(N_L,N_C,N_R) = E_L(N_L) + E_C(N_C) + E_R(N_R) \label{eq:eigenenergies}
\end{equation}
with
\begin{eqnarray}
  E_L(N_L) & = & N_L \left( E_0 + V_b \right)
  + \frac{N_L(N_L - 1)}{2} E_U , \label{eq:EL} \\
  E_C(N_C) & = & N_C \left( E_1 - V_g \right)
  + \frac{N_C(N_C - 1)}{2} E_U ,  \label{eq:EC} \\
  E_R(N_R) & = & N_R \left( E_0 - V_b \right)
  + \frac{N_R(N_R - 1)}{2} E_U . \label{eq:ER}
\end{eqnarray}

Tunneling across the barriers in the potential gives rise to hopping matrix elements
between adjacent single-particle states which are, however, generally much smaller
than the above energy scales,
Their influence on the many-body eigenstates can therefore be neglected, except
for accidental near-degeneracies between the unperturbed energies
(\ref{eq:eigenenergies}) where they give rise to hybridizations between the
involved Fock states $|N_L,N_C,N_R\rangle$.
The resulting avoided crossings in the many-body spectrum are most significant
if the Fock states that participate at this crossing can be mapped into each
other by the exchange of only one atom across one tunneling barrier.

In order to identify the location of these crossings in the parameter space spanned
by the gate and bias voltages, we introduce the local particle-addition and -removal
energies \cite{CapO07PRL} as
\begin{eqnarray}
\mu_{L/C/R}^+(N_{L/C/R}) & \equiv & E_{L/C/R}(N_{L/C/R} + 1) - E_{L/C/R}(N_{L/C/R}) , \\
\mu_{L/C/R}^-(N_{L/C/R}) & \equiv & E_{L/C/R}(N_{L/C/R}) - E_{L/C/R}(N_{L/C/R} - 1) \\
 & = & \mu_{L/C/R}^+(N_{L/C/R} - 1) \nonumber
\end{eqnarray}
for the left, central, and right well, respectively.
With the help of Eqs.~(\ref{eq:EL}--\ref{eq:ER}), we find
\begin{eqnarray}
\mu_L^+(N_L) & = & E_0 + V_b + N_L E_U , \\
\mu_C^+(N_C) & = & E_1 - V_g + N_C E_U , \\
\mu_R^+(N_R) & = & E_0 - V_b + N_R E_U .
\end{eqnarray}
Degeneracies between unperturbed levels that correspond to the populations $N_L$:$N_C$:$N_R$
and $N_L\pm1$:$N_C\mp1$:$N_R$ therefore occur if $\mu_L^\pm(N_L) = \mu_C^\mp(N_C)$, which is
equivalent to the equation
\begin{equation}
  V_b = E_1 - E_0 - V_g + (N_C - N_L \mp 1 ) E_U . \label{eq:VbVgp}
\end{equation}
Correspondingly, we find
\begin{equation}
  - V_b = E_1 - E_0 - V_g + (N_C - N_R \pm 1 ) E_U \label{eq:VbVgm}
\end{equation}
as equation for the degeneracy between $N_L$:$N_C$:$N_R$ and $N_L$:$N_C\pm1$:$N_R\mp1$.

\begin{figure}[t]
\centerline{\includegraphics[width=\linewidth]{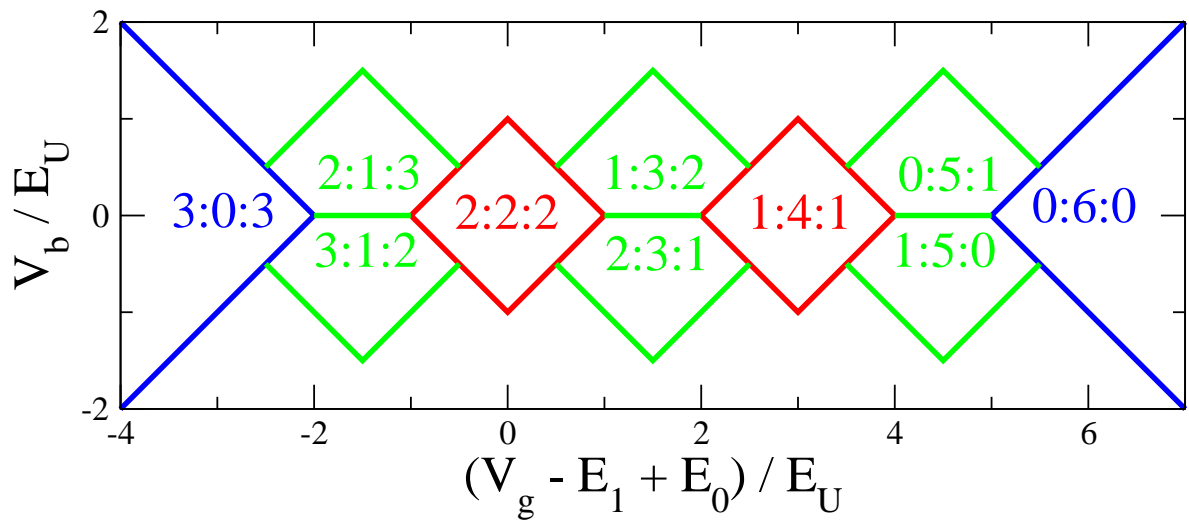}}
\caption{(color online)
  Structure of the interaction-blockade diamonds that result from a simple Bose-Hubbard-type
  model.
  Plotted are the predicted transition lines for single-particle transfer between adjacent
  wells according to Eqs.~(\ref{eq:VbVgp}) and (\ref{eq:VbVgm}), assuming the presence of
  for six particles in a triple-well potential with equal on-site interaction energies.
  The agreement with Fig.~\ref{fig:diamonds} is rather good, taking into account the
  approximate value $E_U \simeq 0.268 V_0$ (\ref{eq:EU}) for the interaction energy.
\label{fig:model}}
\end{figure}

With these equations (\ref{eq:VbVgp}) and (\ref{eq:VbVgm}), we can quantitatively predict the
location and shape of the ``atom blockade diamonds'', i.e.\ of the transition lines that
mark the value of the bias at which single-atom transfer takes place between adjacent wells.
This is shown in the Fig.~\ref{fig:model} for the case of six particles in the triple-well
potential under consideration.
We see that the agreement with the numerically computed transition lines in
Fig.~\ref{fig:diamonds} is rather good.
Taking into account the approximate value (\ref{eq:EU}) for the interaction energy $E_U$,
we can quantitatively reproduce not only the shape, but also the size of the diamond
structures in Fig.~\ref{fig:diamonds}.

There are some important differences to standard Coulomb diamonds in quantum dots.
On the one hand, the diamond structures display ``open ends'' both for very low and for
very high gate voltages, corresponding to an empty quantum dot and to empty
reservoirs, repectively. 
On the other hand, there is a significant asymmetry between the diamond structures
corresponding to a ``balanced'' (with $N_L = N_R$) and an ``unbalanced'' population
(with $N_L = N_R \pm 1$) in the outer wells.
This is a consequence of the finite interaction energy $E_U$ in the reservoirs.

\begin{figure}[t]
\centerline{\includegraphics[width=\linewidth]{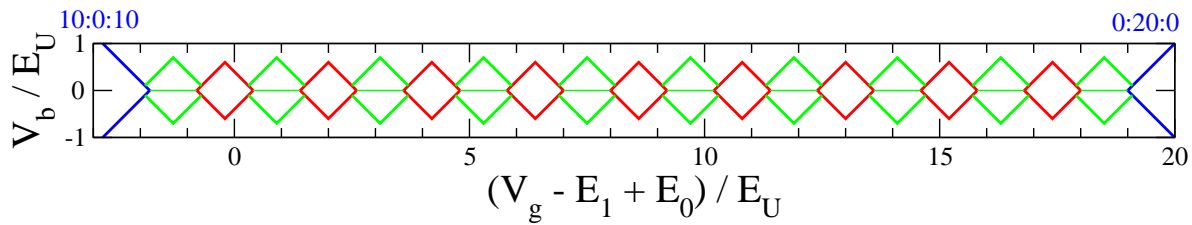}}
\caption{(color online)
  Same as Figure \ref{fig:model} for a more mesoscopic situation with 20 particles,
  where the two outer wells are assumed to be much more shallow and
  thereby exhibit a lower on-site interaction energy, namely $E_U^{(L/R)} = 0.2 E_U^{(C)}
  \equiv 0.2 E_U$, than the central well.
  This situation resembles much more the standard scenario of Coulomb blockade in electronic
  quantum dots where the two-particle interaction energy in the leads does not play a role.
\label{fig:meso}}
\end{figure}

To demonstrate this, we show in Fig.~\ref{fig:meso} a more mesoscopic situation involving
20 particles in total, where the two outer wells are assumed to be much more shallow and
thereby exhibit a lower on-site interaction energy $E_U^{(L/R)}$ than the central well.
This leads to slight modifications of the formulas (\ref{eq:VbVgp}) and (\ref{eq:VbVgm});
we obtain
\begin{eqnarray}
  V_b & = & E_1 - E_0 - V_g + (N_C - 1)E_U^{(C)} - N_L E_U^{(L)} ,  \label{eq:VbVgpmod} \\
  V_b & = & E_1 - E_0 - V_g + N_C E_U^{(C)} - (N_L - 1 ) E_U^{(L)} ,  \label{eq:VbVgpmod2} \\
  - V_b & = & E_1 - E_0 - V_g + (N_C - 1 ) E_U^{(C)} - N_R E_U^{(R)} , \label{eq:VbVgmmod} \\
  - V_b & = & E_1 - E_0 - V_g + N_C E_U^{(C)} - (N_R - 1 ) E_U^{(R)} . \label{eq:VbVgmmod2}
\end{eqnarray}
Specifically assuming $E_U^{(L/R)} = 0.2 E_U^{(C)}$ in Fig.~\ref{fig:meso},
the resulting structure of the transition lines for single-particle transfer
resembles much more the standard Coulomb diamonds in electronic quantum dots.

Independently of the size of the outer wells, the distance between the edges of the
diamonds with balanced populations is, as seen from 
both in Fig.~\ref{fig:model}
and Fig.~\ref{fig:meso}, predicted to be equal to the local interaction energy
$E_U^{(C)}$ in the central well.
This opens the possibility to extract the value of this interaction energy from the
location of the transition lines of single-atom transfer in the $V_g$--$V_b$ parameter
space.
From Fig.~\ref{fig:diamonds} we would approximately infer
$E_U^{(C)} \simeq 0.29 V_0 \simeq 5.8 \hbar^2 k^2 / m$ from the distance between the
corners of the 3:0:3 and the 2:2:2 diamonds.
This is in fairly good agreement with the numerically computed interaction energy
$E_U^{(C)} \simeq 5.1  \hbar^2 k^2 / m$, which is obtained by substracting twice the
lowest single-particle eigenenergy $E_1$ of the central well from the energy of the
lowest two-particle state where both electrons are localized in the central well.

\subsection{Energy scales of the avoided crossings}

While this simple Bose-Hubbard-type model is capable of reproducing the location of the
transition lines for single-atom transfer in the $V_g$--$V_b$ parameter space, it is
\emph{not} sufficient to predict the time scales on which the ramping process ought
to take place in order to achieve those transitions.
As pointed out above, these time scales crucially depend on the sizes of the avoided
crossings in the many-body spectrum, which are essentially given by the single-particle
hopping matrix elements between different wells.
These hopping matrix elements, however, arise from a tunneling process across the
barriers that separate the wells from each other.
They are therefore sensitively depending on the effective \emph{height} of the barriers
with respect to the particle-addition or -removal energies under consideration,
which in turn is appreciably modified under variation of the bias or gate voltages.

Specifically, we find that an increase of the gate voltage, corresponding to ``pulling
down'' the central well with respect to the outer ones, generally gives rise to a
significant enlargement of the avoided crossings between different many-body levels.
This is illustrated in Fig.~\ref{fig:diamonds} in which the size of the circles indicate
the extent $\Delta V_b$ of the avoided crossings between the states under consideration
in bias voltage space.
More precisely, if we describe the general dependence of the level splitting between two
near-degenerate states on the bias voltage by
\begin{equation}
  \Delta E(V_b) = \Delta E \sqrt{ 1 + \left( \nu (V_b - V_b^{(c)}) / \Delta E \right)^2}
\end{equation}
where
$\Delta E \equiv \Delta E(0)$ and $V_b^{(c)}$ denote the splitting and bias voltage,
respectively, at the crossing point and $\nu$ represents the (nearly integer) difference
between the variations of the unperturbed levels with $V_b$ far away from the anticrossing.
We define
\begin{equation}
  \Delta V_b \equiv 2 \Delta E / \nu \label{eq:crossing}
\end{equation}
as the maximal extent of the avoided crossing in bias voltage space.
For anticrossings that correspond to single-atom transfer between adjacent wells, we have
$\nu \simeq 1$ and therefore $\Delta V_b \simeq 2 \Delta E$.

\begin{table}
\caption{
  Level splittings (in units of $\hbar^2 k^2 / m$) of the first two avoided crossings
  that are encountered during the ramping process of the bias voltage,
  for $V_g =0.25 V_0$, $0.5 V_0$, and $1.0 V_0$.
\label{tab:splittings}}
\begin{indented}
\item[]
\begin{tabular}{cc|ccc|ccc}
  & & \multicolumn{3}{c|}{first crossing} & \multicolumn{3}{c}{second crossing} \\
  \hline
  gate & initial & bias & crossing & level &
  bias & crossing & level \\[-1mm]
  voltage & state & voltage & state & splitting & voltage & state & splitting \\
  \hline
  $0.25 V_0$ & 3:0:3 & $0.117 V_0$ & 2:0:4 & 0.0015 & $0.143 V_0$ & 2:1:3 & 0.069 \\
  $0.5 V_0$ & 2:1:3 & $0.248 V_0$ & 1:1:4 & 0.019 & $0.325 V_0$ & 2:0:4 & 0.16 \\
  $1.0 V_0$ & 2:2:2 & $0.128 V_0$ & 1:2:3 & 0.047 & $0.221 V_0$ & 1:3:2 & 0.22
\end{tabular}
\end{indented}
\end{table}

Table \ref{tab:splittings} lists the level splittings at the avoided crossings that
the system undergoes during the time-dependent ramping process, for the gate voltages
$V_g = 0.25 V_0$, $0.5 V_0$, and $1.0 V_0$.
For each of those gate voltages, the first one of the encountered anticrossings
corresponds to a double-barrier tunneling process of an atom which is directly
transferred from the left to the right well.
Correspondingly, the associated anticrossing is much smaller than the one of the
second anticrossing describing a single-barrier tunneling process.
This particularly implies that it is, for all of these cases, possible to define,
according to the Landau-Zener formula (\ref{eq:LZ}), a reasonably large range of
speeds $s$ for the ramping process $V_b(t) = s t$, such that the system undergoes
a diabatic transition across the first and an adiabatic transition at the second
anticrossing.
Comparing, however, the level splittings for \emph{different} values of $V_g$, we
realize that this range of speeds will depend on the particular gate voltage under
consideration, and that it is hardly possible to choose one ``universal'' ramping speed
with which the desired single-atom transfer can be carried out for all gate voltages.
This is another specification of the microscopic nature of this transport system.

The size of the hopping matrix elements do also play a role for the question
to which extent the interaction-blockade phenomena discussed here are also observable
for lower effective one-dimensional interaction strengths $g$, i.e.\ in the presence
of a weaker transverse confinement of the waveguide.
Figure \ref{fig:model} seems to affirm this, as the structure of the diamonds does not
explicitly depend on the interaction strength, except for the scaling of the horizontal
and vertical axes which is directly proportional to $g$.
However, the size and extent of the relevant avoided crossings, on the other hand, does,
in lowest order, \emph{not} scale with $g$, as they are essentially given by the
single-particle hopping matrix element between adjacent sites.
This implies that decreasing the effective interaction strength $g$, e.g., by a factor ten
will, in lowest order, give rise to the same relative location of the transition lines in
Fig.~\ref{fig:diamonds}, with the $V_g$ and $V_b$ axes scaled down by a factor ten, while
the ``uncertainty'' of the transition lines, given by the extent of the corresponding avoided
crossings and represented by the size of the circles in Fig.~\ref{fig:diamonds}, will be
increased by a factor ten.
The interaction-blockade diamonds ultimately become unobservable if the on-site interaction
energy is of the same order as the tunneling matrix element between adjacent sites, in which
case \emph{collective} instead of single-atom transfer processes begin to take place.
This underlines again the obvious requirement that the transport processes discussed here
have to be performed in the ``Mott-insulator regime'' of strong on-site interaction and weak
tunneling within the triple-well potential.

\section{Discussion and conclusion}

\label{sec:conc}

In summary, we studied a microscopic analog of source-drain transport with ultracold
bosonic atoms in a triple-well potential.
The latter is considered to be realized by optical triple-well lattices in analogy with
the experiments of Refs.~\cite{FoeO07N,CheO08PRL}, which are initially loaded with a
well-defined number of atoms per site.
We propose to perform then a time-dependent tilt of these triple-well potentials until a
given maximal bias, followed by measuring the resulting populations in the individual wells.
This process can be repeated for various values of a ``gate voltage'' which lowers the level
of the central well with respect to the outer ones.
Diamond structures are obtained for the transition lines that mark the necessary values for
the bias in order to achieve single-atom transfer between adjacent wells, when being plotted
as a function of that gate voltage.
These diamonds can be used to infer the local interaction energy in the central well.

There are striking similarities with Coulomb diamonds in electronic quantum dots, but also
some important differences to the latter, which mainly arise due to the microscopic nature
of the ``reservoirs'' in the outer two wells.
Most characteristically, ``transport'' is manifested not by a continuos flow of particles
across the quantum dot, but rather by the transfer of a single atom from one well to another.
This transport process is, in general, not ``completed'' insofar as the atom when, e.g.,
being transferred from the left to the central well will not directly go on to the right well,
due to the mismatch of the corresponding particle-addition and -removal energies.
Close to the corners of the diamond structures (e.g.\ at $V_g \simeq 0.3 V_0$, see
Fig.~\ref{fig:diamonds}), this latter task can eventually be achieved by further
increasing the bias.
In general, however, more complicated ramping processes, involving also a time-dependent
variation of the gate voltage, would be required for that purpose.
This is clearly a consequence of the finite two-body interaction energy in the reservoirs,
which also leads to a striking asymmetry between diamond structures corresponding to balanced
and unbalanced populations in the outer wells.

The microscopic transport scheme that we investigated here can be applied as well to
more complex atomic gases involving, e.g., several spin components and/or long-range
dipolar interaction.
This will open various possibilities for the exploration of the interplay of interaction
and transport in an experimentally feasible context.
Our findings should, moreover, be relevant for 
for cyclic processes~\cite{HilKotCoh08EPL}
as well as for the evaluation of the feasibility of
atomtronic scenarios and atomic transistors \cite{SeaO07PRA,StiAndZoz07PRA,PepO09PRL}
and might provide valuable insight for the design of logical operations with single atoms.

\ack

We would like to thank I. Bloch, I. B\v{r}ezinov\'a, K. Capelle, 
G. Kavoulakis, M. Oberthaler, C. Pethick,
M. Rontani, A. Wacker, and S. Z\"ollner for useful discussions.
This work was financed by the Swedish Research Council and the
Swedish Foundation for Strategic Research.
The collaboration is part of the NordForsk 
Nordic network {\it Coherent Quantum Gases - From
Cold Atoms to Condensed Matter}
\section*{References}


\begin{thebibliography}{10}
\bibitem{note}
We subdivide for this purpose the time interval $\delta t$ into 10000
  sub-intervals within which we assume the Hamiltonian to be locally
  time-independent. The propagation within such a sub-interval is performed by
  explicit matrix exponentiation.
\end{thebibliography}

\end{document}